\documentclass[final,authoryear,10pt,twocolumn,twoside]{elsarticle}

\usepackage{amssymb}
\usepackage{graphicx}
\usepackage{textcomp}
\usepackage{amsmath}
\usepackage{setspace}

\begin{document}

\begin{frontmatter}

\title{Detecting chaos in a complex system}

\author{Boyan Hristozov Petkov}
\address{Institute of Atmospheric Sciences and Climate
   (ISAC) of the Italian National Research Council (CNR), Via Gobetti 101,
    I-40129 Bologna, Italy, e-mal:b.petkov@isac.cnr.it}

\begin{abstract}
The sequences, given by a 7D map have been analysed by means of the
methods, widely used to detect chaos in the real world in order to
test their sensitivity to chaotic features of a non-linear system
determined by comparatively high number of parameters. The same
diagnostic approaches have been applied to the 3D Lorenz map for
comparison. The results show that for some of the sequences yielded
from the 7D map, the adopted methods were not able to give as
straightforward answer to the question if the system is chaotic as
in the 3D case. Since the sequences, subject of the analysis, were
not contaminated by noise and were sufficiently long, it could be
assumed that such difficulties have arisen likely due to specific
internal features of the more complex system. It was found also that
an increase from 0.01 to 0.5 of the sampling step determining the
sequences obtained from the 7D map, masks the chaos in some of them.

\end{abstract}

\end{frontmatter}


\section{\label{sec1}Introduction}
Very often, the decision about chaotic origin of a time series turns
out to be a fundamental issue for the study of the processes in the
real world \citep{Lorenz1991,Marzocchi1997,Lai2002,Voss2009}. Such a
judgement is usually made basing on the assessment of the attractor
invariants -- Housdorff dimension and Lyapunov exponents
\citep{Marzocchi1997,Lai2002,Grassberger1991,Kodba2005}. Their
correct estimation provides important information about the
complexity of the system and its dynamics that helps us to create an
adequate model of the phenomenon under study.

The reliability of the methods, developed for detecting chaos in the
real systems is usually tested on time series yielded from well
studied maps. A similar test has been performed here by applying the
adopted methods to the sequences obtained from a 7-dimensional (7D)
map, defined by 7 parameters, all connected through nonlinear
relationships. Such a map was assumed to imitate a comparatively
more complex system and, in order to highlight eventual its specific
features, the same test has been done with the sequences given by a
3D map. Except for the invariants, the minimal dimension of the
space, embedding the reconstructed attractor has been also
evaluated, since this parameter gives the number of variables
composing the system that is also an important characteristic.

To study the sensitivity of the methods that compute the invariants,
the researchers usually take one component of a map assuming that
the others should represent the same behaviour
\citep{Lorenz1991,GrassbergerProcaccia1984,Zeng1991}. In the present
study, these methods have been applied to each of the sequences,
yielded from both 3D and 7D maps. Such a performance was adopted in
order to check if each of the sequences generated by the 7D map is
able to depicter adequately the system attractor. It was assumed
also that the sampling step determining the sequences under study is
able to influence the topological properties of the reconstructed
attractor and hence, the assessment of the parameters chosen to
characterise the system. To analyse such an assumption the
evaluations of the corresponding parameters were made by varying the
sampling steps of the sequences under study.

\section {\label{sec2}Methods of nonlinear time series analysis used in the present study}
A starting point for the analysis of a nonlinear system presented by
a time series is the reconstruction of the embedding phase space and
the attractor. Furthermore, an assessment of the Hausdorff dimension
$D_{0}$ approximated by estimators allowing easy computation and
Lyapunov exponents $\lambda_{l}\,(l=1,2,..,m)$ is usually performed.
Next subsections briefly present some widely used algorithms for
estimation of these parameters, which have been applied in the
present analysis.

\subsection{\label{sec2.1}Reconstruction of the system attractor}
The reconstruction of the attractor using only one scalar projection
\citep{Packard1980,Takens1981} gave a powerful instrument to study
the natural phenomena. For a sequence $(x_{i})_{i=1,\,2,\,..,\,N}$
determined by measuring of the variable $x$ at uniquely sampled
times $(t_{i})_{i=1,\,2,\,..,\,N}$ the Takens's theorem
\citep{Takens1981} affirms that the $m-$component vectors
constructed as
\begin{equation}\label{Eq1}
\mathbf{X}_{i}=(x_{i},\,x_{i+\tau},\,x_{i+2\tau},...,\,x_{i+(m-1)\tau})
\, ,
\end{equation}
where $\tau$ is the so-called time delay, expressed in sampling
steps $\Delta t=t_{i+1}-t_{i}$, determine a manifold that
realistically represents the attractor of the system, which
generates the time series $(x_{i})$. It should be pointed out that
there is no a unique optimal choice of the time delay $\tau$
\citep{Grassberger1991,Zeng1992}. In the present study the parameter
$\tau$ is taken to be the time for which the autocorrelation
function drops to $e^{-1}$ or to about 0.37.

\subsection{\label{sec2.2}Minimum dimension of the reconstructed embedding space}
According to \citet{Kennel1992} an acceptable minimum embedding
dimension $m_{K}$ of the attractor can be assessed by looking at the
behaviour of the nearest neighbor $\mathbf{X}^{(b)}$ of each vector
$\mathbf{X}_{i}$ when the embedding dimension $m$ increases.
Assuming Euclidean metric in the phase space the authors found that
each $\mathbf{X}^{(b)}$ can be considered as a false nearest
neighbor if either of the following two conditions:
\begin{equation}
\label{Eq2}
\frac{|x_{i+m\tau}-x_{i+m\tau}^{(b)}|}{\|\mathbf{X}_{i}-\mathbf{X}^{(b)}\|_{(m)}^{(E)}}>R_{tol}
\end{equation}\\
and
\begin{equation}
\label{Eq3}
\frac{\|\mathbf{X}_{i}-\mathbf{X}^{(b)}\|_{(m+1)}^{(\text{E})}}{R_{A}}>A_{tol}\,,
\end{equation}\\
is held. The expression
$\|\mathbf{X}_{i}-\mathbf{X}^{(b)}\|_{(m)}^{(\text{E})}=
\sqrt{\sum_{k=1}^{m}\bigl(x_{i+(k-1)\tau}-x^{(b)}_{i+(k-1)\tau}\bigr)^{2}}$
denotes the Euclidean distance between two vectors in
$m-$dimensional embedding spaces and according to \citet{Kennel1992}
$R_{tol}$ can be considered higher than 10 and $A_{tol}=2$. The
parameter $R_{A}$ represents the size of the attractor and was taken
to be equal to the standard deviation of
$(x_{i})_{i=1,\,2,\,..,\,N}$. \citet{Kennel1992} assumed that the
minimum dimension of the embedding space $m_{K}$ for which the false
nearest neighbors percentage (FNNP) drops to a value below 1\%,
allows unfolding of the attractor. Studding the 3D Lorenz system
they found also that for a noise free sequence the FNNP remains
lower than 1\% for $m>m_{K}=3$, while a noise contaminated time
series shows a different behaviour. For low level of the noise the
approach gave $m_{K}=4$, whereas for higher level FNNP falls to a
value slightly exceeding 1\% at $m=m_{K}$ and plateaus for higher
embedding dimension. In presence of strong noise the length of such
a plateau narrows to a few successive values of $m$ and after that
FNNP increases. In case of sequence presenting a stochastic process
FNNP drops to a comparatively high value ($>20\%$) and after that
rapidly increases.

\subsection{\label{sec2.3}Correlation dimension of the attractor}
Correlation dimension $D_{2}$ of the attractor, which is a widely
used estimator of the Hausdorff dimension $D_{0}$, can be assessed
by calculating the correlation integral $C_{m}(\rho)$:
\begin{eqnarray}
\label{Eq4} C_{m}(\rho)\,=\, \lim_{N \to \infty}
\frac{2}{(N+1-W)(N-W)}\nonumber\times\\
\times\sum_{j=W}^{N}\sum_{i=1}^{N-j} \theta(\rho- \Vert
\mathbf{X}_{i}-\mathbf{X}_{i+j} \Vert^{(\text{Ch})}_{(m)})\,,
\end{eqnarray}\\
where $\theta (\xi)$ is the Heaviside function $\bigl(\theta (\xi<0)
= 0$ and $\theta (\xi \geq 0) = 1\bigr)$ and
$\|\mathbf{X}_{p}-\mathbf{X}_{q}\|^{(\text{Ch})}_{(m)}=
\underset{1\leq k\leq
m}{\max}\,\bigl\{|x_{p+(k-1)\tau}-x_{q+(k-1)\tau}|\bigr\}$ is the
distance between two vectors in $m$-dimensional embedding space
determined here by the Chebishev metric. The correlation integral
$C_{m}(\rho)$ was defined by \citet{GP1983} as Eq (\ref{Eq4}) gives
it for $W=1$ and later, \citet{Theiler1986} proposed the
introduction of the cutoff parameter $W$ to avoid a spurious
estimate of the correlation dimension resulted from high
autocorrelation in the time series under study.

The main point of the \citet{GP1983} analysis was the affirmation
that for small $\rho$ the correlation integral scales as a power of
$\rho$:
\begin{equation}
\label{Eq5} C_{m}(\rho)\, \sim \, \rho^{D_{m}}\,.
\end{equation}
Determining the parameter  $D_{m}$ by averaging the local slopes:
\begin{equation}
\label{Eq6} D_{m}(\rho) \,=\, \frac{\Delta \bigl[\text{lg}\bigl(
C_{m}(\rho)\bigr)\bigr]}{\Delta \bigl(\lg(\rho)\bigr)}
\end{equation}\\
over $\bigl\{\rho:D_{m}(\rho)=\text{const}\bigr\}$, the correlation
dimension can be assessed as $D_{2}= \underset{m \to \infty} \lim
D_{m}$ \citep{Lorenz1991,LaiLerner1998}. In case of chaotic system
the sequence $D_{m}$ rapidly increases to its limit and the minimal
value of the embedding dimension $m_{GP}$ for which $D_{m}$ reaches
a plateau, presents another estimator of the minimum embedding
dimension for the attractor \citep{Cao1997,LaiLerner1998}. The
analysed time series can be considered as being resulted from a
chaotic system if the correlation dimension $D_{2}$ is a small
fractal number, whereas for a stochastic sequence $\underset{m \to
\infty} \lim D_{m}=\infty\,$.

Various studies discussed the minimum sampling size of a time series
needed for the correct estimation of the correlation dimension
$D_{2}$ \citep{Ruelle1990,NerenbergEssex1990,Theiler1990}. However,
\citet{Grassberger1991} argued against the existence of an optimal
time series length, affirming that such a claim could take place for
other generalized dimensions but not for $D_{2}$.

\subsection{\label{sec2.4}Lyapunov spectrum and Kaplan-Yorke dimension of the attractor}
Lyapunov exponents characterise the divergence of the orbits in the
attractor and they were determined here following the approach
proposed by \citet{Eckmann1985} and further developed by others
\citep{Eckmann1986,Zeng1991,Zeng1992,Zeng1992_1}. The method traces
out the growth of the distances between a vector $\mathbf{X}_{i}$
and each of the vectors $\mathbf{X}_{j}$ for which
$\varepsilon_{min}<\|\mathbf{X}_{j}-\mathbf{X}_{i}\|_{(m)}^{(\text{E})}<\varepsilon$,
where $\varepsilon_{min}$ and $\varepsilon$ are small numbers. The
evolution of these differences, over $\tau$ steps ahead on the
fiducial trajectory can be determined by a matrix
$T_{i}:\,\mathbf{X}_{j+\tau}-\mathbf{X}_{i+\tau}=T_{i}(\mathbf{X}_{j}-\mathbf{X}_{i})$,
which is computed for all consecutive vectors $\mathbf{X}_{i}$
($i=1,1+\tau,1+2\tau,...,K$), where
$K\leq\bigl(N-(m-1)\tau-1\bigr)/\tau$. Furthermore, the matrices
$T_{i}$ ($i=1,2,...,K$) are successively reorthogonalized by means
of a standard $Q_{i}R_{i}$ decomposition \citep{Eckmann1985} and the
Lyapunov exponents are given by \citep{Eckmann1986,Zeng1991}
\begin{equation}
\label{Eq7} \lambda_{l}=\frac{1}{\tau
K}\sum_{i=1}^{K}\ln(R_{i})_{ll}\,.
\end{equation}
Taking natural logarithm, the above equation gives the Lyapunov
exponents $\lambda_{l}$ in $\bigl($bits/(sampling
step)$\bigr)\cdot$ln2. Finally, the exponents $\lambda_{l}$ are
evaluated as averages of the corresponding values found by varying
the parameters $\varepsilon_{min},\,\varepsilon$ and $m$. The method
determines the exponents $\lambda_{l}$ in order
$\lambda_{1}>...>\lambda_{l}>...>\lambda_{m}$ and one of them should
be identified as zero. According to \citet{Zeng1992_1} the ability
of the method to estimate correctly the negative exponents is
limited. A chaotic system is characterised by at least one positive
$\lambda_{l}$ and as higher the positive Lyapunov exponents are as
faster is the orbit divergence in the attractor that makes the
correct prediction of the future states less reliable even in the
case of negligible errors in the initial conditions.

\citet{KaplanYorke1978} introduced another estimator $D_{KY}$ of the
Hausdorff dimension $D_{0}$ defined as:
\begin{equation}
\label{Eq8}
D_{KY}=k+\frac{\sum_{l=1}^{k}\lambda_{l}}{|\lambda_{k+1}|}\,,
\end{equation}
where $k=\underset{1\leq l\leq
m}{\max}\,(l:\sum_{l=1}^{k}\lambda_{l}\geq0)$ and both estimators
$D_{2}$ and $D_{KY}$ are related to $D_{0}$ according to
\citep{GP1983,Farmer1983}:
\begin{equation}
\label{Eq9} D_{2}\leq D_{0} \leq D_{KY}\,.
\end{equation}

\citet{EckmannRuelle1992} claimed that the correlation dimension
$D_{2}$ and Lyapunov exponents can be correctly estimated if the
sampling size $N$ of the time series satisfy the inequality:
\begin{equation}
\label{Eq9_1} Q\lg N\geq D_{2}\,,
\end{equation}
where $Q=2$ if we deal with the correlation dimension $D_{2}$ and
$Q=1$ when the Lyapunov exponents should be evaluated.

\subsection{\label{sec2.5}Surrogate data}
\citet{Theiler1992} proposed a test for nonlinearity in time series
based on the construction of surrogate data from the sequence under
study. A widely used approach to creating surrogates applies the
Fourier transform to the data and after the randomization of the
phases of the obtained spectral components, the inverse Fourier
transform returns the surrogate time series, which has the same
statistical properties as the original one. Under the null
hypothesis that the analysed sequence is stochastic, the estimates
of the above parameters found for the surrogate data, confirm or
reject this hypothesis if they coincide with or differ from those
obtained for the original time series \citep{Theiler1992}.

\begin{figure}[h]
\centerline{\includegraphics{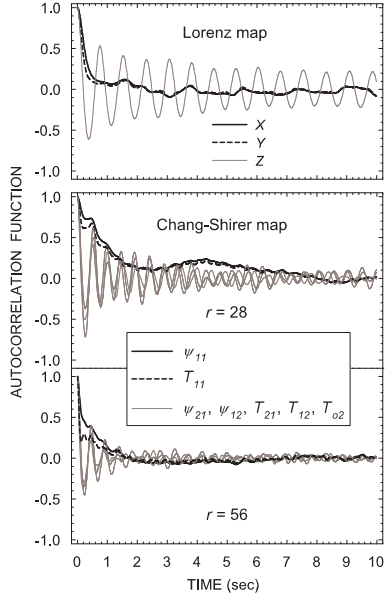}}
\caption{\label{F1}Autocorrelation functions of the sequences under
study arbitrary assuming the sampling step $\Delta t$ as being
expressed in seconds. The upper part exhibits the autocorrelation in
the three solutions of the Lorenz map, while the lower part shows
the autocorrelation in each of the seven sequences yielded from the
Chang-Shirer map in both $r=28$ and $r=56$ cases.}
\end{figure}

\begin{figure}[h]
\centerline{\includegraphics{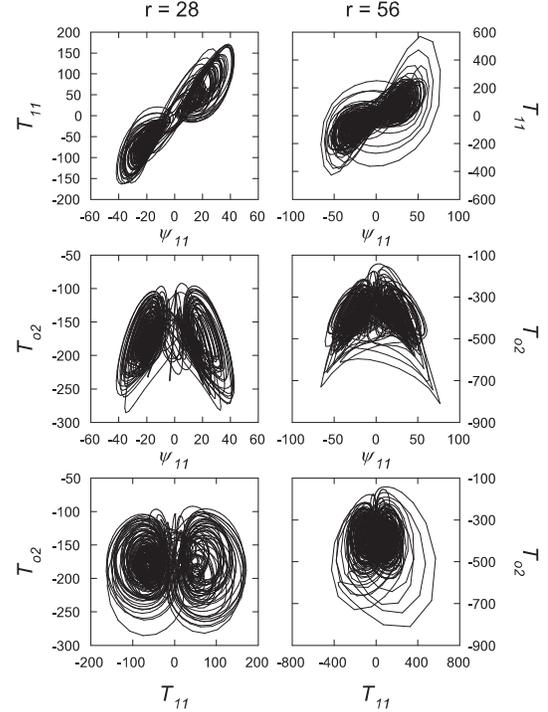}} \caption{\label{F2}Some
projections of the Chang-Shirer attractors corresponding to $r=28$
on the left and $r=56$, on the right in case of $\Delta t=0.01$.}
\end{figure}

\section {\label{sec3}Sequences used in the present analysis}
The methods, shortly described in the previous section are commonly
used to judge whether a time series represents one-dimensional
projection of a chaotic system or not. To test their sensitivity to
detect chaos in a complex system, they were applied to each of the
sequences obtained as solutions of a 7D map taken to mimic a system
characterised by a higher extent of complexity. Simultaneously, a 3D
map considered an example of a less complex system  was a subject of
the same studies for comparison. The well known Lorenz map
\citep{Lorenz1963} defined as:
\begin{eqnarray}\label{Eq10}
  \dot{X} &=& \sigma (Y-X)\nonumber\\
  \dot{Y} &=& X(\rho-Z)-Y \\
  \dot{Z} &=& XY-\beta Z\nonumber
\end{eqnarray}
where $\sigma=10,\,\rho=28$ and $\beta=8/3$, has been chosen to
represent this case. The former system was presented by a map
defined by \citet{ChangShirer1984} as (hereinafter referred to as
the Chang-Shirer map or system):
\begin{eqnarray}\label{Eq11}
  \dot{\psi}_{m1} = aC_{5}C_{7}\psi_{n1}\psi_{(m-n)2}/4C_{1}\nonumber\\
  +PmaT_{m1}/C1-PC_{1}\psi_{m1}\nonumber\\
  \dot{\psi}_{n1} = -aC_{5}C_{6}\psi_{m1}\psi_{(m-n)2}/4C_{2}\nonumber\\
  +PnaT_{n1}/C_{2}-PC _{2}\psi_{n1}\nonumber\\
  \dot{\psi}_{(m-n)2} = -a^{3}C_{4}C_{5}^{2}\psi_{m1}\psi_{n1}/4C_{3}\nonumber\\
  +PaC_{4}T_{(m-n)2}/C_{3}-PC_{3}\psi_{(m-n)2}\nonumber\\
  \dot{T}_{m1} = aC_{5}[\psi_{n1}T_{(m-n)2}-\psi_{(m-n)2}T_{n1}]/4\\
  +ma\psi_{m1}T_{o2}+maR\psi_{m1}-C_{1}T_{m1}\nonumber\\
  \dot{T}_{n1} = aC_{5}[\psi_{m1}T_{(m-n)2}+\psi_{(m-n)2}T_{m1}]/4\nonumber\\
  +na\psi_{n1}T_{o2}+naR\psi_{n1}-C_{2}T_{n1}\nonumber\\
  \dot{T}_{(m-n)2} = -aC_{5}[\psi_{m1}T_{n1}+\psi_{n1}T_{m1}]/4\nonumber\\
  +aC_{4}R\psi_{(m-n)2}-C_{3}T_{(m-n)2}\nonumber\\
  \dot{T}_{o2} =
  -a[m\psi_{m1}T_{m1}+n\psi_{n1}T_{n1}]/2-4T_{o2}\nonumber\,,
\end{eqnarray}
where
$C_{1}=1+m^{2}a^{2},\,C_{2}=1+n^{2}a^{2},\,C_{3}=4+(m-n)^{2}a^{2},\,C_{4}=m-n,\,
 C_{5}=m+n,\,C_{6}=3+C_{4}^{2}a^{2}-m^{2}a^{2}$ and $C_{7}=3+C_{4}^{2}a^{2}-n^{2}a^{2}$.
Analysing this system, \citet{Nese1987} found that for
$m=2,\,n=1,\,a=\sqrt{2}/2,\,P=10$ and $R=6.75\,r$ it provides
chaotic solutions.

The three sequences given by the Lorenz map and two groups of
sequences obtained from the Chang-Shirer map \citep{Nese1987} for
(i) $r=28$ and initial conditions (100, -150, -100, -1200, -2000,
-5000, -2500), and (ii) $r=56$ with initial vector (-6, 20, -15, 10,
5, 15, 16) were obtained with sampling steps $\Delta t=0.01,\,
0.05,\,0.1\,\, \text{and}\,\,0.5$, respectively. All these solutions
have been found by using the ``ode45'' MATLAB procedure for 20000
sampling steps, omitting the first 10000 to avoid transitions. The
sampling size of $N=10000$ was taken to satisfy inequality
(\ref{Eq9_1}) (see also Table \ref{T1}) despite some objections  to
the claim about the minimal sampling size \citep{Grassberger1991}.

\begin{figure*}
\centerline{\includegraphics{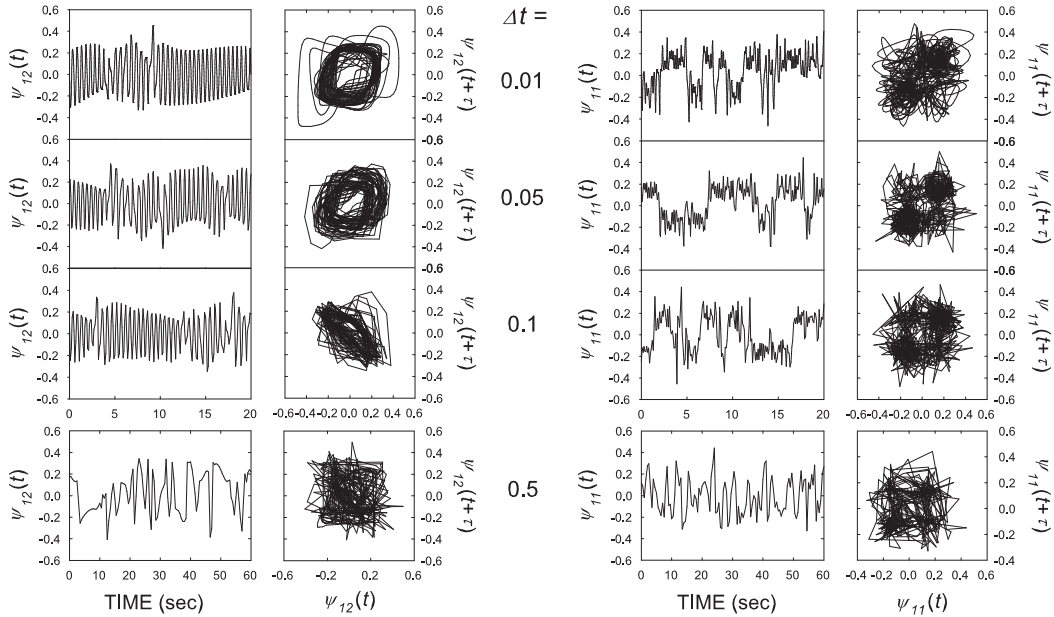}} \caption{\label{F3}The two
columns on the left present the time series and 2D projections of
the corresponding attractors reconstructed from $\psi_{12}(r=28)$
for different steps $\Delta t$ indicated in the central part of the
figure. The right two columns exhibit the same for
$\psi_{11}(r=56)$.}
\end{figure*}

\begin{table}[h]\footnotesize \caption{\label{T1} The three Lyapunov exponents
of the Lorenz map and the largest five ones computed for the
Chang-Shirer system for $r=28$ and $r=56$, respectively together
with the Kaplan-Yorke ($D_{KY}$) and correlation ($D_{2}$)
dimensions of the attractors, all evaluated by \citet{Nese1987} and
considered reference values in the present analysis. The Lyapunov
exponents are given in $\bigl($(bits/sec)$\cdot$ln2$\bigr)$.}
\begin{tabular}{p{0.12in}p{0.14in}p{0.2in}cccc}\\
\hline\\
 $\lambda_{1}$&$\lambda_{2}$&$\lambda_{3}$&$\lambda_{4}$&$\lambda_{5}$&$D_{KY}$&$D_{2}$\\
 \\ \hline\\
 \multicolumn{7}{c}{Lorenz map}\\\\
 $0.93$&$0.0$&$-14.60$&---&---&$2.063$&$2.05\pm0.01$\\\\
 \multicolumn{7}{c}{Chang-Shirer map, $r=28$}\\\\
 $0.84$&$0.07$&$0.00$&$-0.43$&$-18.60$&$4.03$&$2.9\pm0.1$\\\\
 \multicolumn{7}{c}{Chang-Shirer map, $r=56$}\\\\
 $2.51$&$1.12$&$0.00$&$-1.80$&$-16.30$&$4.11$&$3.8\pm0.1$\\\\
 \hline
 \end{tabular}
\end{table}

Figure \ref{F1} presents the autocorrelation functions for these
three groups of sequences adopted for the present analysis. It is
seen that the time series presented by $\psi_{11}$ in both $r=28$
and $r=56$ cases and, $T_{11}$ for $r=28$ are characterised by
comparatively high autocorrelation.

Some 2D projections of the Chang-Shirer attractors constructed for
$r=28$ and $r=56$ respectively, are shown in Fig. \ref{F2}, while
Fig. \ref{F3} exhibits the reconstructed attractors from
$\psi_{12}(r=28)$ and $\psi_{11}(r=56)$ corresponding to different
sampling steps $\Delta t$. Each of the sequences under study has
been normalized by $\bigl(\max (x_{i})-\min (x_{i})\bigr)_{1\leq i
\leq N}$ that limits the differences $|x_{i}-x_{j}|_{1\leq i,j \leq
N}$ between 0 and 1, and facilitate the calculation of the
parameters used in the analysis. Figure 3 shows that the time series
patterns found for $\Delta t=0.05$ and 0.1 do not differ
significantly from those at $\Delta t=0.01$ in both cases shown on
the left and right, while for $\Delta t=0.5$ the corresponding
sequences look completely different. However, the corresponding 2D
phase portraits indicate that the reconstructed attractors are more
sensitive to the variations in the sampling time $\Delta t$. It is
clearly seen that only for $\Delta t=0.01$ the projections of the
reconstructed attractors are depicted by smooth curves like those
presented in Fig. \ref{F2}. For lower resolution (higher $\Delta t$)
the attractors turn out to be represented by broken-line orbits and
for $\Delta t=0.5$ the fiducial trajectory is composed in practice
by long segments. A similar loss of the typical features of the
attractor, resulted from enlarging of the sampling time, was
reported by \citet{Kim_Yoon2001} who analysed the Lorenz map. Thus,
it can be concluded that an attractor projection depicted by
broken-line orbits acts as an indicator for large sampling step in
the sequence under study. However, it should be pointed out that the
noise is able to produce a similar effect \citep{Kawata1997}.

\begin{table}[h]\footnotesize
\caption{\label{T2}Minimum embedding dimension $m_{K}$ found for
each of the attractors reconstructed using the sequences under
study, determined as $m_{K}=\min\,(m:\text{FNNP}<1\%)$. The values
found in case when FNNP follows the noise contaminated pattern (see
Fig. \ref{F4} and the text), determined as
$m_{K}=\min\,(m:\text{FNNP}<n\%)$ are signed by an asterisk together
with the lowest value $n\%$ of FNNP, if it exceeds 1\%, given in
parenthesis. The cases for which FNNP behaves similarly to a
stochastic sequence shown in Fig. \ref{F4} are signed by ``--''. The
values of $m_{GP}$ are also given, separated from $m_{K}$ by
semicolons; ``$\infty$'' indicates that the corresponding value can
not be determined varying $m$ from 1 to 20.}\ \tabcolsep=0.11cm
\centering \begin{tabular}{p{0.28in}cccc} \hline
 \\$\Delta t=$&0.01&0.05&0.1&0.5\\
 \\\hline\\
 \multicolumn{5}{c}{Lorenz map}\\\\
 $X$&3;\,3&3;\,3&3;\,3&3;\,3\\
 $Y$&3;\,3&3;\,3&3;\,3&3;\,3\\
 $Z$&3;\,3&3;\,3&3;\,3&4;\,3\\\\
 \multicolumn{5}{c}{Chang--Shirer map, r=28}\\\\
 $\psi_{21}$&4;\,7&5;\,7&$4^{\ast}$(2\%);\,7&--;$\infty$\\
 $\psi_{11}$&4;\,$\infty$&5;\,7&5;\,4&--;$\infty$\\
 $\psi_{12}$&4;\,4&4;\,5&4;\,5&4;\,7\\
 $T_{21}$&4;\,7&4;\,7&4;\,7&4;$\infty$\\
 $T_{11}$&6,$\infty$&$11^{\ast}$(2\%);\,$\infty$&4;\,6&4;\,6\\
 $T_{12}$&4;\,4&$4^{\ast}$;\,4&$4^{\ast}$;\,5&--;\,6\\
 $T_{o2}$&4;\,4&4;\,5&4;\,5&4;$\infty$\\\\
 \multicolumn{5}{c}{Chang--Shirer map, r=56}\\\\
 $\psi_{21}$&5;\,7&$5^{\ast}$(3\%);\,7&$5^{\ast}$(3\%);\,7&--;$\infty$\\
 $\psi_{11}$&6;\,$\infty$&4;$\infty$&--;$\infty$&--;$\infty$\\
 $\psi_{12}$&7;\,5&4;\,7&4;\,6&4;$\infty$\\
 $T_{21}$&4;\,6&4;\,6&4;\,6&4;$\infty$\\
 $T_{11}$&5;\,7&4;\,7&4;\,7&4;$\infty$\\
 $T_{12}$&4;\,6&4;\,7&4;\,7&4;$\infty$\\
 $T_{o2}$&10;\,7&$4^{\ast}$(5\%);\,6&$4^{\ast}$(5\%);\,6&4;$\infty$\\\\
 \hline
 \end{tabular}
\end{table}

Table \ref{T1} gives the Lyapunov spectra, Kaplan-Yorke ($D_{KY}$)
and correlation ($D_{2}$) dimensions estimated by \citet{Nese1987}
for the attractors corresponding to Lorenz and two cases of the
Chang-Shirer systems, which values have been used here as reference
ones.

\section{\label{sec4}Results and discussion}
\begin{figure}
\centerline{\includegraphics{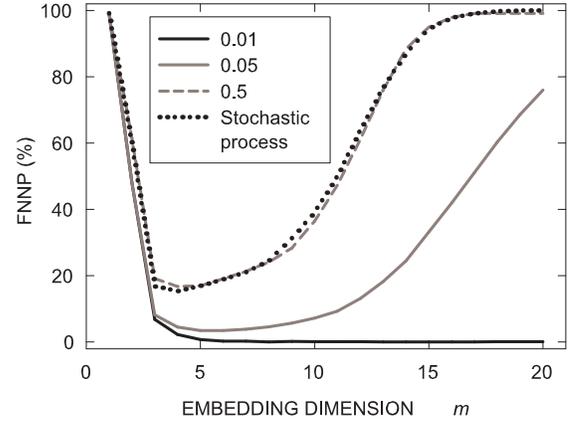}} \caption{\label{F4}The false
nearest neighbors percentage (FNNP) found as a function of embedding
dimension for $\psi_{21}(r=56)$ in case of three different sampling
times $\Delta t$ equal to 0.01, 0.05 and 0.5, respectively. The FNNP
determined for a sequens of random values with Gaussian
distribution, considered to represent a stochastic process is also
given for comparison. In all cases the parameter $R_{tol}$ was
assumed to be equal to 30.}
\end{figure}

Table \ref{T2} exhibits the minimum embedding dimensions $m_{K}$ and
$m_{GP}$ of the attractors, reconstructed from the sequences under
study. As can be seen, for the Lorenz attractor the FNNP approach
gave an estimation of $m_{K}$ equal to the real embedding dimension
$m_{L}=3$, with one exception slightly exceeding this value.
Conversely, the estimates found for the two groups of the time
series obtained from Chang-Shirer system in case of $r=28$ and
$r=56$ respectively, showed values varying between 4 and 11. Figure
\ref{F4} represents the behaviour of FNNP as a function of the
embedding dimension for $\psi_{21}(r=56)$ obtained for different
time resolutions. The curve corresponding to the sequence with
sampling time of 0.01 drops to 0.73\% at $m=5$ and remains below
this value until embedding dimension increases up to 20. A similar
behaviour showed FNNP in all cases of the time series $X$, $Y$ and
$Z$ obtained from the Lorenz system. The FNNP curve found for
$\psi_{21}(r=56,\Delta t=0.05)$ follows a comportment similar to
that presented by noise contaminated time series, while the curve
characterising $\psi_{21}(r=56,\Delta t=0.5)$ shows a behaviour
typical for the attractor reconstructed from a stochastic time
series. Table \ref{T2} shows that both $m_{K}$ and $m_{GP}$
estimators give in practice equal values for the attractors of the
Lorenz system, except for $Z(\Delta t=0.5)$ where a slight
difference between them is seen. However, the application of the
same approaches to the sequences yielded from the Chang-Shirer
system do not give such an accordance between $m_{K}$ and $m_{GP}$
estimates. It is seen that $m_{K}$ tends to be underestimated and
only in case of $\psi_{12}(r=56,\Delta t=0.01)$, $m_{K}$ represents
the actual embedding dimension $m_{CS}=7$. In contrast,
$m_{GP}=m_{CS}$ for 32\% of all the sequences obtained from the
Chang-Shirer map for the adopted values of $\Delta t$ but for 30\%
of them the corresponding approach does not give an estimate, while
such a percentage is 11 \% for the FNNP method.

\begin{figure}[h]
\centerline{\includegraphics{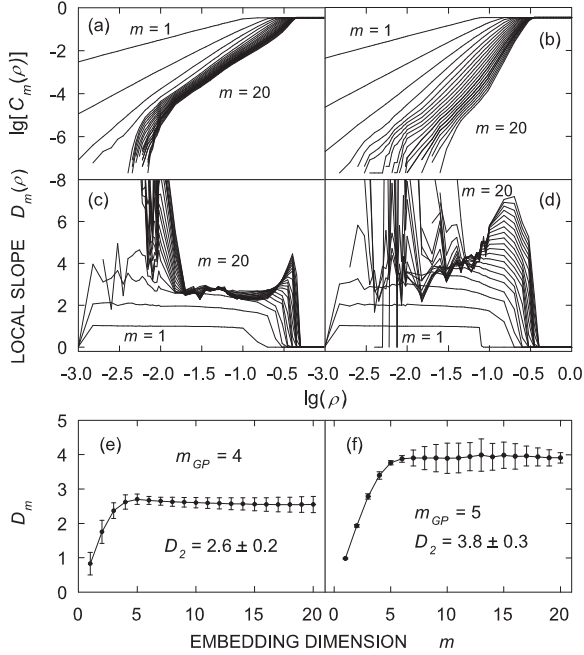}} \caption{\label{F5}
Behaviour of the correlation integral $C_{m}(\rho)$ as a function of
$\rho$ presented in decimal logarithm scale (a, b) and the
corresponding local slope $D_{m}(\rho)$ (c, d) evaluated for the
sequence $\psi_{12}(r=28,\,\Delta t=0.01)$ (left) and
$\psi_{12}(r=56,\,\Delta t=0.01)$ (right). In both cases the
embedding dimension $m$ gradually increased from 1 to 20. Panels (e,
f) present the behaviour of the parameter $D_{m}$ as a function of
$m$ and the corresponding values of $m_{GP}$ and $D_{2}$ are also
indicated.}
\end{figure}

\begin{figure}
\centerline{\includegraphics{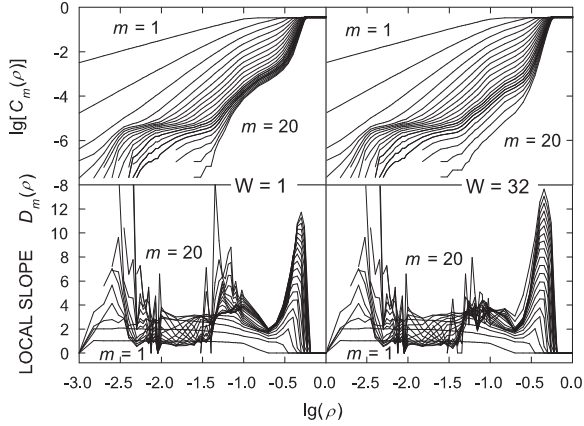}} \caption{\label{F6}Scaling
behaviour of the correlation integral $C_{m}(\rho)$ as a function of
$\rho$ in decimal logarithm scale (upper panels) and the
corresponding local slope $D_{m}(\rho)$ (lower panels) assessed for
the time-series $\psi_{11}(r=28,\,\Delta t=0.05)$ for two different
values of parameter $W$ (see Eq. \ref{Eq4}).}
\end{figure}

Figure \ref{F5} shows as an example the scaling behaviour of the
correlation integral $C(\rho)$ as a function of $\rho$ in decimal
logarithm scale and the corresponding variations in the local slop
$D_{m}(\rho)$ for $\psi_{12}(\Delta t=0.01)$ obtained assuming
$r=28$ and $r=56$, respectively. The lower part of Fig. \ref{F5}
illustrates the behaviour of parameter $D_{m}$ defined from the
curves $D_{m}(\rho)$ $\bigl($see Eq. (\ref{Eq6})$\bigl)$. It can be
seen that both cases presented in Fig. \ref{F5} exhibit different
scaling patterns of the correlation integral. Panels (c) and (d)
indicate that the linear part is easy recognizable for $r=28$, while
for $r=56$ it becomes shorter and less marked. Figure \ref{F6}
illustrates the comportment of the correlation integral evaluated
for $\psi_{11}(r=28,\,\Delta t=0.05)$, one of the sequences
characterised by high autocorrelation (see Fig. \ref{F1}). Assuming
$W=1$, Eqs. (\ref{Eq4}) and (\ref{Eq6}) give the curves shown on the
left-hand side of Fig. \ref{F6}, while the curves corresponding to
$W=\tau=32$ can be seen on the right. The figure indicates that a
linear segment of $\lg\bigl(C(\rho)\bigr)$ can not be identified for
$W=1$, while taking $W=32$ a very short plateau in the local slope
could be recognized for $-1.3<\lg(\rho)<-0.8$. However, despite the
use of the cutoff parameter $W$ the behaviour of the correlation
integral slightly changes that makes the estimation of the
correlation dimension to be on the edge of the reliability. In
contrast, the three series yielded by the Lorenz system for the
adopted values of $\Delta t$ show a comparatively long linear
segment in the corresponding curves similarly to the case given in
Figs. \ref{F5} (a) and (c).

\begin{table}[h]\footnotesize
\caption{\label{T3}Estimators of the Hausdorff dimension $D_{0}$ of
the attractors under study presented by the corresponding
correlation dimension $D_{2}$ with its standard deviation $\Delta
D_{2}$ and the Kaplan-Yorke dimension $D_{KY}$ both shown as
``$D_{2}\pm\Delta D_{2};\,D_{KY}$''. The symbol ``$\infty$''
indicates that a finite value of $D_{2}$ has not been found for
embedding dimension increasing up to 20.} \ \tabcolsep=0.11cm
\begin{tabular}  {p{0.12in}cccc} \hline
 \\$\Delta t=$&0.01&0.05&0.1&0.5\\
 \\ \hline\\
 \multicolumn{5}{c}{Lorenz map}\\\\
 $X$&2.07$\pm$0.09;\,2.3&2.06$\pm$0.04;\,2.4&2.05$\pm$0.08;\,2.5&2.05$\pm$0.08;\,2.9\\
 $Y$&2.08$\pm$0.09;\,2.3&2.04$\pm$0.09;\,2.3&2.05$\pm$0.05;\,2.2&2.05$\pm$0.08;\,2.9\\
 $Z$&2.10$\pm$0.08;\,2.1&2.08$\pm$0.06;\,2.2&2.17$\pm$0.07;\,2.4&2.10$\pm$0.08;\,2.6\\\\
 \multicolumn{5}{c}{Chang--Shirer map, r=28}\\\\
 $\psi_{21}$&2.7$\pm$0.2;\,4.8&3.1$\pm$0.2;\,4.6&3.2$\pm$0.1;\,4.6&$\infty$;\,5.1\\
 $\psi_{11}$&$\infty$;\,4.8&3.3$\pm$0.6;\,4.3&2.3$\pm$0.4;\,5.3&$\infty$;\,4.6\\
 $\psi_{12}$&2.6$\pm$0.2;\,4.6&3.2$\pm$0.2;\,4.3&3.3$\pm$0.2;\,4.6&3.7$\pm$0.2;\,4.6\\
 $T_{21}$&2.6$\pm$0.1;\,4.6&3.2$\pm$0.2;\,4.6&3.2$\pm$0.2;\,4.9&$\infty$;\,4.6\\
 $T_{11}$&$\infty$;\,4.4&$\infty$;\,4.3&3.5$\pm$0.5;\,4.4&3.7$\pm$0.6;\,4.6\\
 $T_{12}$&2.6$\pm$0.1;\,4.6&3.0$\pm$0.3;\,4.5&3.3$\pm$0.2;\,4.5&3.7$\pm$0.2;\,4.7\\
 $T_{o2}$&2.8$\pm$0.1;\,4.4&3.3$\pm$0.1;\,4.8&3.3$\pm$0.1;\,4.4&$\infty$;\,4.4\\\\
 \multicolumn{5}{c}{Chang--Shirer map, r=56}\\\\
 $\psi_{21}$&3.9$\pm$0.2;\,4.4&3.7$\pm$0.2;\,4.7&3.9$\pm$0.1;1,4.8&$\infty$;\,4.1\\
 $\psi_{11}$&$\infty$;\,5.1&$\infty$;\,4.5&$\infty$;\,3.0&$\infty$;\,3.6\\
 $\psi_{12}$&3.8$\pm$0.3;\,4.5&3.9$\pm$0.2;\,4.5&3.9$\pm$0.1;\,4.4&$\infty$;\,3.4\\
 $T_{21}$&3.9$\pm$0.1;\,4.6&3.8$\pm$0.3;\,4.6&3.9$\pm$0.1;\,4.6&$\infty$;\,3.5\\
 $T_{11}$&3.7$\pm$0.1;\,4.5&3.8$\pm$0.1;\,4.6&4.0$\pm$0.1;\,4.6&$\infty$;\,3.7\\
 $T_{12}$&3.8$\pm$0.4;\,4.7&3.8$\pm$0.1;\,4.3&3.9$\pm$0.1;\,4.3&$\infty$;\,3.6\\
 $T_{o2}$&3.9$\pm$0.2;\,4.5&4.1$\pm$0.2;\,4.5&4.0$\pm$0.1;\,4.6&$\infty$;\,3.5\\\\
 \hline
 \end{tabular}
\end{table}

The behaviour of the correlation integral presented in the upper
part of Fig. \ref{F5} reveals an interesting feature. The increasing
slope at low $\rho$ significantly more pronounced for $r=28$ lead to
assume the presence of noise in the corresponding attractors
\citep{Theiler1990,Eckmann1985} even though such a component was not
added solving Eqs. (\ref{Eq11}). A similar conclusion can be made
analysing the behaviour of FNNP as a function of the embedding
dimension $m$ for some of the sequences yielded from the
Chang-Shirer map as Fig. \ref{F4} and Table \ref{T2} show. It should
be pointed out that an analogous occurrence was not observed
studding the 3D Lorenz system.

Table \ref{T3} exhibits the correlation dimension $D_{2}$ of the
attractors reconstructed from the time series under study. It can be
seen that the values of $D_{2}$ found for the sequences of the
Lorenz system for all adopted sampling steps are very close to the
reference ones given in Table \ref{T1} . A similar behaviour shows
the correlation dimensions $D_{2}$ of the Chang-Shirer attractors
reconstructed for $r=56$ and $\Delta t=0.01,\,0.05$ and 0.1. For
$r=28$ Table \ref{T3} exhibits a slight increase of $D_{2}$ when
$\Delta t$ increases. Surprisingly, Table \ref{T3} indicates also
that some of the series provided by the Chang-Shirer system
determine an attractor for which a finite correlation dimension
cannot be found. In case of $r=56$ the parameter $D_{2}$ is infinite
for $\psi_{11}$ found at all adopted $\Delta t$, while for $\Delta
t=0.5$ such an occurrence characterises all the sequences. In case
of $r=28$ the attractors with undefined correlation dimension are
those constructed from $\psi_{11}$ for $\Delta t=0.01$ and 0.5, and
$T_{11}$ for $\Delta t=0.01$ and 0.05. For $\Delta t=0.5$ such
features show also $\psi_{21}$, $T_{21}$ and $T_{02}$. It should be
pointed out that the assessment of the correlation dimension $D_{2}$
for $\psi_{11}(r=28,\,\Delta t=0.1)$ and $T_{11}(r=28,\,\Delta
t=0.1,\,\,\text{and}\,\,0.5)$ was very difficult to make, like in
the case of $\psi_{11}(r=28,\,\Delta t=0.05)$ shown in Fig.
\ref{F6}.

\begin{table}[h] \footnotesize \caption{\label{T4}Lyapunov exponents
of the Lorenz system obtained from the analysis of the corresponding
sequences, arbitrary assuming $\Delta t$ in seconds.}\ \centering
\begin{tabular}{p{0.12in}ccc} \hline
 \\&$\lambda_{1}$&$\lambda_{2}$&$\lambda_{3}$\\
 \\ \hline\\
 \multicolumn{4}{c}{$\Delta t= 0.01$ s}\\\\
 X&0.98$\pm$0.07&0.0&-3.30$\pm$0.60\\
 Y&1.09$\pm$0.04&0.0&-3.20$\pm$0.10\\
 Z&0.94$\pm$0.09&0.0&-6.10$\pm$0.20\\\\
 \multicolumn{4}{c}{$\Delta t= 0.5$ s}\\\\
 X&1.00$\pm$0.10&0.0&-1.10$\pm$0.70\\
 Y&1.12$\pm$0.02&0.0&-1.26$\pm$0.09\\
 Z&0.89$\pm$0.09&0.0&-1.50$\pm$0.60\\\\
 \hline
\end{tabular}
\end{table}

Tables \ref{T4} and \ref{T5} represent the Lyapunov spectra of the
sequences obtained from the Lorenz and Chang-Shirer sistems,
respectively. As can be seen the positive Lyapunov exponents
evaluated for the attractors of the Lorenz system in case of $\Delta
t=0.01$ and 0.5 are in good agreement with the reference values
given in Table \ref{T1}. Similar results, not shown in Table
\ref{T4}, were found for 0.05 and 0.1 sampling steps, as well. Table
\ref{T5} exhibits the five largest Lyapunov exponents found for all
the seven sequences yielded from the Chang-Shirer system at $\Delta
t=0.01$ and 0.5  for both $r=28$ and $r=56$ values. In case of
$r=28$ and $\Delta t=0.01$ the positive exponents turn out to be
overestimated except for the sequences $\psi_{11}$ and $T_{11}$. The
same features present the time series found for 0.05 and 0.1
sampling times (not shown in Table \ref{T5}). In contrary, for
$\Delta t=0.5$ the exponent $\lambda_{1}$ turned out to be
underestimated. Except for $\psi_{11}$, the positive values of
$\lambda_{l}$ have been correctly estimated for all the sequences
obtained for $r=56$ at sampling times $\Delta t=0.01,\,0.05$ and 0.1
(the last two cases are not shown in Table \ref{T5}), while for
$\Delta t=0.5$ the assessments of the Lyapunov exponents give
completely different results characterised by one, appreciably
underestimated positive exponent.

\begin{table}[h] \tiny \caption{\label{T5}The first five Lyapunov exponents
evaluated from the Chang-Shirer sequences for both adopted values of
parameter $r$ and sampling steps of 0.01 and 0.5, arbitrary assuming
$\Delta t$ in seconds.}\ \tabcolsep=0.24cm \begin{tabular}
{p{0.12in}ccccc}
\hline\\
 &$\lambda_{1}$&$\lambda_{2}$&$\lambda_{3}$&$\lambda_{4}$&$\lambda_{5}$\\\\
 \hline\\
 \multicolumn{6}{c}{r=28}\\\\
 \hline\\
 \multicolumn{6}{c}{$\Delta t= 0.01$ s}\\\\
 $\psi_{21}$&1.27$\pm$0.06&0.43$\pm$0.06&0.0&-0.60$\pm$0.20&-1.40$\pm$0.40\\
 $\psi_{11}$&0.51$\pm$0.03&0.15$\pm$0.02&0.0&-0.15$\pm$0.01&-0.62$\pm$0.05\\
 $\psi_{12}$&1.40$\pm$0.10&0.60$\pm$0.10&0.0&-0.65$\pm$0.02&-2.40$\pm$0.10\\
 $T_{21}$&1.30$\pm$0.09&0.55$\pm$0.03&0.0&-0.70$\pm$0.10&-2.00$\pm$0.20\\
 $T_{11}$&0.36$\pm$0.07&0.22$\pm$0.06&0.0&-0.31$\pm$0.02&-0.71$\pm$0.03\\
 $T_{12}$&1.40$\pm$0.10&0.70$\pm$0.10&0.0&-0.75$\pm$0.09&-2.10$\pm$0.10\\
 $T_{o2}$&1.22$\pm$0.06&0.60$\pm$0.10&0.0&-0.80$\pm$0.10&-2.81$\pm$0.03\\\\
 \multicolumn{6}{c}{$\Delta t= 0.5$ s}\\\\
 $\psi_{21}$&0.19$\pm$0.06&0.06$\pm$0.02&0.0&-0.07$\pm$0.01&-0.14$\pm$0.02\\
 $\psi_{11}$&0.18$\pm$0.03&0.07$\pm$0.02&0.0&-0.12$\pm$0.02&-0.23$\pm$0.01\\
 $\psi_{12}$&0.26$\pm$0.05&0.11$\pm$0.02&0.0&-0.16$\pm$0.02&-0.35$\pm$0.01\\
 $T_{21}$&0.13$\pm$0.03&0.03$\pm$0.01&0.0&-0.07$\pm$0.01&-0.15$\pm$0.01\\
 $T_{11}$&0.18$\pm$0.03&0.07$\pm$0.01&0.0&-0.11$\pm$0.02&-0.21$\pm$0.02\\
 $T_{12}$&0.26$\pm$0.09&0.12$\pm$0.06&0.0&-0.13$\pm$0.02&-0.33$\pm$0.03\\
 $T_{o2}$&0.16$\pm$0.04&0.05$\pm$0.02&0.0&-0.13$\pm$0.02&-0.22$\pm$0.01\\\\
 \hline\\
 \multicolumn{6}{c}{r=56}\\\\
 \hline\\
 \multicolumn{6}{c}{$\Delta t= 0.01$ s}\\\\
 $\psi_{21}$&2.60$\pm$0.20&1.70$\pm$0.10&0.0&-2.30$\pm$0.50&-4.50$\pm$0.30\\
 $\psi_{11}$&0.59$\pm$0.07&0.33$\pm$0.04&0.0&-0.24$\pm$0.04&-0.55$\pm$0.08\\
 $\psi_{12}$&2.60$\pm$0.20&1.51$\pm$0.04&0.0&-2.10$\pm$0.10&-4.30$\pm$0.10\\
 $T_{21}$&2.70$\pm$0.30&1.40$\pm$0.30&0.0&-1.26$\pm$0.09&-4.50$\pm$0.50\\
 $T_{11}$&2.60$\pm$0.20&1.50$\pm$0.10&0.0&-2.00$\pm$0.50&-4.00$\pm$0.30\\
 $T_{12}$&2.20$\pm$0.40&1.10$\pm$0.30&0.0&-1.20$\pm$0.20&-2.90$\pm$0.20\\
 $T_{o2}$&2.70$\pm$0.30&1.20$\pm$0.10&0.0&-1.60$\pm$0.10&-4.50$\pm$0.30\\\\
 \multicolumn{6}{c}{$\Delta t= 0.5$ s}\\\\
 $\psi_{21}$&0.60$\pm$0.20&0.0&-0.17$\pm$0.01&-0.38$\pm$0.05&-1.00$\pm$0.40\\
 $\psi_{11}$&0.58$\pm$0.09&0.0&-0.21$\pm$0.04&-0.60$\pm$0.04&-1.43$\pm$0.01\\
 $\psi_{12}$&0.50$\pm$0.10&0.0&-0.24$\pm$0.01&-0.64$\pm$0.03&-1.39$\pm$0.07\\
 $T_{21}$&0.54$\pm$0.07&0.0&-0.24$\pm$0.02&-0.55$\pm$0.09&-1.42$\pm$0.03\\
 $T_{11}$&0.50$\pm$0.10&0.0&-0.15$\pm$0.02&-0.48$\pm$0.03&-1.38$\pm$0.06\\
 $T_{12}$&0.24$\pm$0.01&0.0&-0.09$\pm$0.01&-0.26$\pm$0.02&-0.67$\pm$0.05\\
 $T_{o2}$&0.49$\pm$0.08&0.0&-0.22$\pm$0.03&-0.59$\pm$0.04&-1.37$\pm$0.03\\\\
 \hline
  \end{tabular}
\end{table}

In all cases the last component of the Lyapunov spectra turned out
to be significantly overestimated due to the limited ability of the
method to estimate correctly the negative exponents
\citep{Zeng1992_1}. Such an overestimation caused a corresponding
overestimation of the Kaplan-Yorke dimension $D_{KY}$ (see Table
\ref{T1} and Table \ref{T3}). For the Lorenz system, the parameter
$D_{KY}$ was found to be higher by 2\%--40\% with respect to the
reference values given in Table \ref{T1}, while such an amount for
the Chang-Shirer system varied from 9\% to 20\% for $r=28$ and from
5\% to 14\% for $r=56$, respectively. On the other hand, the
dimension $D_{KY}$ has been determined in practice for each of the
attractors and, when the correlation dimension $D_{2}$ exists, the
relationship between them, expressed by inequalities (\ref{F9}) is
always held. In case of the Chang-Shirer system, except for the
sequences at $r=56$ and $\Delta t=0.5$, and for
$\psi_{11}(r=56,\,\Delta t=0.1)$, all the estimates of $D_{KY}$ are
very similar to each other even in the cases when the Lyapunov
exponents were not correctly evaluated. Thus, it can be concluded
that $D_{KY}$ is slightly sensitive to the internal features of the
system that could impact the other estimators analysed here, on the
one hand, and to the variations in the sampling step, on the other.

\begin{figure}
\centerline{\includegraphics{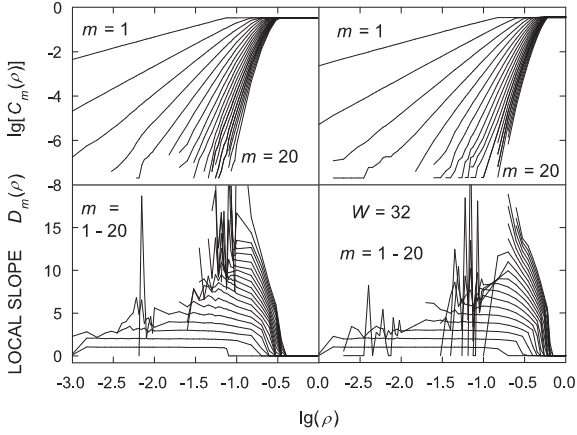}} \caption{\label{F7} As in
Fig. \ref{F6} but for surrogates of $\psi_{12}(r=56,\,\Delta
t=0.01)$ (left) and $\psi_{11}(r=28,\,\Delta t=0.05)$ (right).}
\end{figure}

To have more clear idea about the characteristics of the attractors
reconstructed from the time series under study, the corresponding
surrogate sequences were created as was described in Section
\ref{sec2.5}. Figure \ref{F7} illustrates the behaviour of the
correlation integral calculated for surrogates corresponding to
$\psi_{12}(r=56,\,\Delta t=0.01)$ and $\psi_{11}(r=28,\,\Delta
t=0.05)$. Comparing the curves of Fig. \ref{F7} with those presented
in Fig. \ref{F5} (upper right part) and Fig. \ref{F6} (right)
respectively, that concern the original sequences, it can be seen
that the surrogates exhibit quite different patterns. Hence, despite
of the hardly recognizable linear segment in the curves
$C_{m}(\rho)$ corresponding to the original sequences
$\psi_{12}(r=56,\,\Delta t=0.01)$ and $\psi_{11}(r=28,\,\Delta
t=0.05)$ it can be conclude that we deal with time series provided
by a chaotic system.

\begin{figure}[h]
\centerline{\includegraphics{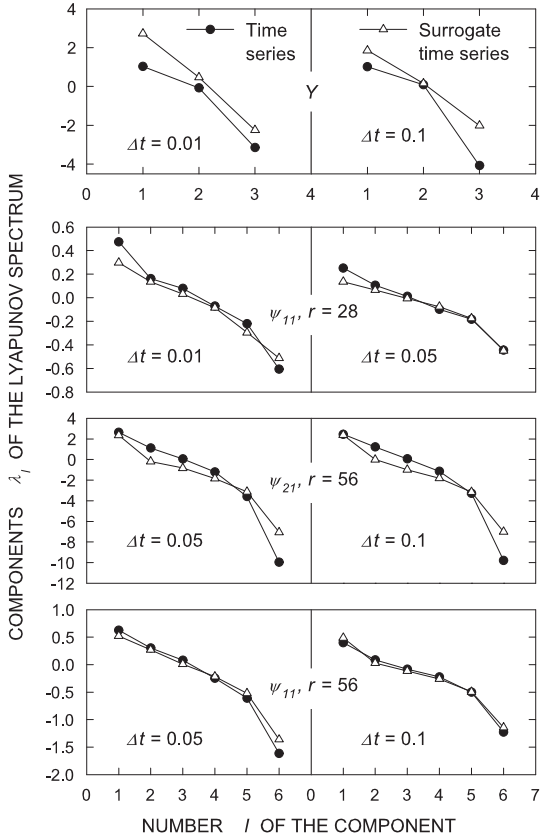}} \caption{\label{F8}Lyapunov
exponents for some of the time series determined by the Lorenz and
Chang-Shirer systems together with the values found for the
corresponding surrogate data.}
\end{figure}

\begin{figure}[h]
\centerline{\includegraphics{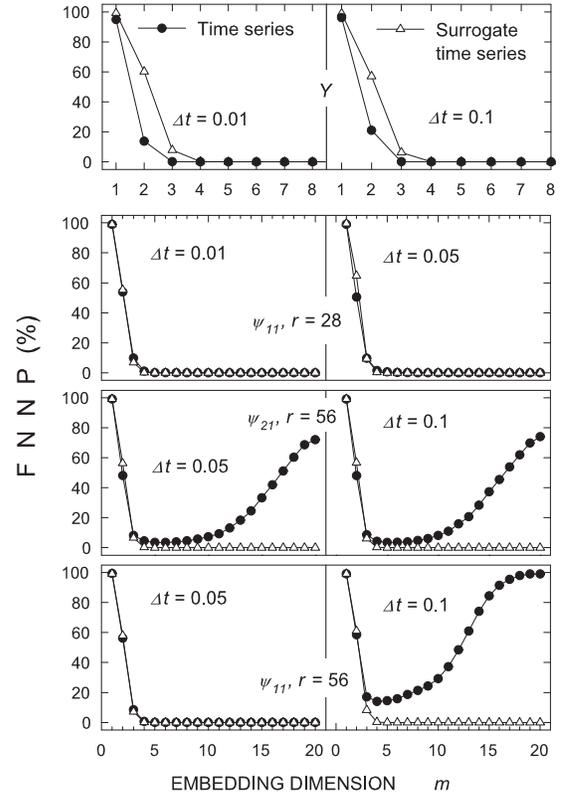}}
\caption{\label{F9}Variations in the false nearest neighbors
percentage (FNNP) for the same as in Fig. (\ref{F8}) sequences.}
\end{figure}

The results presented in Tables \ref{T2}, \ref{T3} and \ref{T5}
identify some of the sequences under study as particular cases. To
understand better the behaviour of these sequences, except for the
correlation dimension $D_{2}$, the Lyapunov spectrum $\lambda_{l}$
and minimum embedding dimension $m_{K}$ of the corresponding
surrogate attractors have been also evaluated. While the parameter
$D_{2}$ showed features similar to those given in Fig. \ref{F7} (not
shown here), the results for $\lambda_{l}$ and $m_{K}$ presented
some particularities. Figures \ref{F8} and \ref{F9} demonstrate the
estimates of these parameters for some cases of $\psi_{11}$ and
$\psi_{21}$ together with the $Y$ component of the Lorenz system.
For the latter, it is seen that the surrogate data present different
patterns of $\lambda_{l}$ and $m_{K}$ respectively, for both $\Delta
t=0.01$ and $\Delta t=0.1$ (see Figs. \ref{F8} and \ref{F9}).
Similarly, the Lyapunov spectra for $\psi_{21}$ sequences and the
corresponding surrogates are different. Despite of the same
embedding dimension, identified as 5 for both original and surrogate
sequences $\psi_{21}(\Delta t=0.05\,\,\text{and}\,\,0.1)$ as Fig.
\ref{F9} shows, the surrogate FNNP behaves similarly to a noise-free
time series, while it follows the noise-contaminated comportment for
the original sequences (see Fig. \ref{F4}). Only the first positive
component of the Lyapunov spectra for $\psi_{11}(r=28)$ and the
corresponding surrogates in both $\Delta t=0.01$ and $\Delta t=0.05$
cases are different (Fig. \ref{Eq8}), while the FNNP approach shows
no differences between $\psi_{11}(r=28)$ sequences and their
surrogates (Fig. \ref{Eq9}). The Lyapunov spectra of
$\psi_{11}(r=56,\,\Delta t=0.05\,\,\text{and}\,\,0.1)$ for both
original and surrogate sequences are almost equal to each other
(Fig. \ref{Eq8}). For surrogate of $\psi_{11}(r=56,\,\Delta
t=0.05)$, FNNP behaves similarly to the original sequence as Fig.
\ref{Eq9} indicates, while the surrogate of $\psi_{11}(r=56,\,\Delta
t=0.1)$ exhibits a behaviour quite different from the corresponding
original sequence. In fact, while for the original data FNNP shows a
typical for stochastic time series comportment, the corresponding
surrogate data present a FNNP behaviour characterising a noise-free
chaotic time series.

Thus, while for the 3D Lorenz map the conclusion about chaotic
character of the system was quite straightforward, for the
Chang-Shirer map some of the sequences ran into difficulties. In
fact, let we assume that a blind test using the sequences yielded
from the second map should be performed. If we make a conclusion
about chaotic origin just on the basis of the estimated correlation
dimension $D_{2}$ we will take the wrong decision attributing
stochastic features to considerable number of the sequences. In
addition for instance, if $\psi_{11}(r=28,\,\Delta t=0.01)$ is the
subject of the analysis, the adopted methods give $m_{K}=4$,
undefined $m_{GP}$, $D_{2}=\infty$, $D_{KY}=4.8$, and two positive
Lyapunov exponents $\lambda_{1}=0.51\pm0.03$ and
$\lambda_{2}=0.15\pm0.02$. The surrogate test shows differences just
in the first Lyapunov exponent and, as a result in $D_{KY}$, which
are $0.29\pm0.03$ and 5.1, respectively. Thus, if a researcher had
at his/her disposal these estimates, he/she would likely conclude
that the system, which generated this series is stochastic. In case
of $\psi_{11}(r=56,\,\Delta t=0.05)$ such a decision would seem more
grounded. Since a noise component was not added solving the analysed
maps and the sampling size of the sequences was chosen to satisfy
the conditions assumed to assure a correct estimation of the
invariants, it could be concluded that the difficulties in detecting
chaos have arisen likely due to specific internal features of the 7D
system. It should be pointed out that the last two examples showing
that the chaotic origin of the corresponding sequences is hardly
recognizable concerned the system components, characterised by high
autocorrelation as Fig. \ref{F1} shows.

The results reported by \citet{Zeng1992} illustrate behaviour
similar to that described in this section. The authors examined the
sequences yielded from surface temperature and pressure
measurements. The analysis showed infinite or unreliably high
correlation dimension of the reconstructed attractors. On the other
hand, it was found that these attractors were characterised by two
positive Lyapunov exponenets and despite that the Kaplan-Yorke
dimension was not evaluated, it is easy to conclude that $D_{KY}$
varies between 4 and 5. Although the question about chaotic origin
of the sequences was not raised by \citet{Zeng1992}, the results of
the present study allow the conclusion that the time series analysed
by them had been likely generated by chaotic processes.

\section{\label{sec5}Conclusions}
The parameters, most commonly used to judge whether a time series
represents one-dimensional projection of a chaotic system have been
estimated for each of the sequences generated by both 3D Lorenz and
7D Chang-Shirer maps considering the second as a more complex
system. The sequences were not contaminated by additional noise and
their sampling sizes were assumed to assure a correct estimation of
the correlation dimension and Lyapunov exponents. In addition, the
impact of the sampling step on the assessed parameters was
investigated.

Performed analysis highlighted some important features of the
reconstructed attractors. First of all, the adopted methods gave an
unambiguous answer to the question if each of the sequences provided
by the 3D Lorenz map has a chaotic origin. Moreover, the estimated
parameters showed very similar values for all sampling times,
assumed here. In contrary, not each of the sequences yielded from
the 7D map represented correctly the attractor properties as in the
case of the 3D system. For some of the sequences of the 7D map the
false nearest neighbors approach together with the correlation
integral behaviour characterised the system as stochastic especially
in the case of larger sampling steps. When these approaches gave an
assessment, the minimum embedding dimension turned out to be
generally underestimated, while the correlation dimension $D_{2}$
was predominantly correctly evaluated. For the 7D sequences
characterised by high autocorrelation, a finite value of $D_{2}$ was
impossible to find in the most of the cases and the corresponding
positive Lyapunov exponents $\lambda_{l}^{+}$ turned out to be
underestimated. A similar estimates of $D_{2}$ and $\lambda_{l}^{+}$
were obtained for the time series with large sampling step. Among
all the evaluated parameters just the Kaplan-Yorke estimator
$D_{KY}$ of the Housdorff dimension of the attractor gave reliable
values for the major part of the sequences yielded from the 7D map.
The surrogate data test applied to some time series did not show
differences between certain parameters evaluated for the original
and the corresponding surrogate sequences, that usually
characterises a stochastic system.

The present study shows that the widely used methods for detecting
chaos in systems of the real world could run into difficulties with
a sequence generated by a high-dimensional process even in case when
it is yielded from a theoretical map without noise contamination and
presenting a sufficient sampling size. Thus, it can be concluded
that the decision about chaotic origin of a time series provided by
an experiment or field observations should be taken with a special
caution, taking into account the estimations of several parameters
that characterise the corresponding attractor. Even if only one or
two of these parameters give a positive answer, the hypothesis about
chaotic origin of the time series should not be excluded.

\textheight=126pt

\footnotesize
\bibliographystyle{model2-names}

\end{document}